# Discovery of H-alpha Emission from the Close Companion Inside the Gap of Transitional Disk HD142527[0]

Close, L.M.[1], Follette, K.B.[1], Males, J.R.[1,6], Puglisi, A.[2], Xompero, M.[2], Apai, D.[1,3], Najita, J.[4], Weinberger, A.J.[5], Morzinski, K.[1,6], Rodigas, T.J.[1], Hinz, P.[1], Bailey, V[1]., Briguglio, R.[2]

lclose@as.arizona.edu

[1]*Steward Observatory, University of Arizona, Tucson, AZ 85721, USA*

[2]*INAF - Osservatorio Astrofisico di Arcetri, I-50125, Firenze, Italy*

[3] *Lunar and Planetary Laboratory, University of Arizona, Tucson, AZ 85721, USA*

[4]*NOAO, 950 N Cherry Ave. Tucson, AZ 85719, United States*

[5]*Carnegie Institution DTM, 5241 Broad Branch Rd, Washington, DC 20015 USA*

[6]*NASA Sagan Fellow*

## ABSTRACT

We utilized the new high-order 585 actuator Magellan Adaptive Optics system (MagAO) to obtain very high-resolution visible light images of HD142527 with MagAO's VisAO science camera. In the median seeing conditions of the 6.5m Magellan telescope (0.5 − 0.7″), we find MagAO delivers 24-19% Strehl at Hα (0.656 μm). We detect a faint companion (HD142527B) embedded in this young transitional disk system at just 86.3±1.9 mas (~12 AU) from the star. The companion is detected in both Hα and a continuum filter (Δmag=6.33±0.20 mag at Hα and 7.50±0.25 mag in the continuum filter). This provides confirmation of the tentative companion discovered by Biller and co-workers with sparse aperture masking at the 8m VLT. The Hα emission from the ~0.25 solar mass companion (EW=180 Angstroms) implies a mass accretion rate of ~5.9x10$^{-10}$ M$_{sun}$/yr, and a total accretion luminosity of 1.2% L$_{sun}$. Assuming a similar accretion rate, we estimate that a 1 Jupiter mass gas giant could have considerably better (50-1000x) planet/star contrasts at Hα than at H band (COND models) for a range of optical extinctions (3.4-0 mag). We suggest that ~0.5-5 M$_{jup}$ extrasolar planets in their gas accretion phase could be much more luminous at Hα than in the NIR. This is the motivation for our new MagAO GAPplanetS survey for extrasolar planets.

[0]This paper includes data gathered with the 6.5 meter Magellan Telescopes located at Las Campanas Observatory, (LCO) Chile.

# 1.0   INTRODUCTION

While it is well established that planet formation is one of the most poorly understood problems in astrophysics today, there has been very little progress made towards the goal of directly imaging very young giant planets during the brief but luminous phase when they are accreting most of their gaseous atmosphere (Fortney et al. 2008; Mordasini et al. 2012; Spiegel & Burrows 2012).

**1.1 Very Brief Overview of Past Work in field: No Images of Hα Accreting Extrasolar Planets**

While direct detection of accreting gas giants at r>5 AU is a key goal of modern direct imaging surveys it has not yet been achieved. Several wide planetary systems have been discovered, notably HR8799bcde (Marois et al. 2008; Skemer et al. 2012) and the hot-start planet Beta-Pic b (Lagrange et al. 2009; Bonnefoy et al. 2013), however these systems lie in debris disks and have passed the gas-accretion phase of formation. Young planetary mass objects that might still be weakly accreting have been imaged. One such planet is 2MASS1207b, a ~4-8 Jupiter mass companion ~50 AU from an 8 Myr old ~25 Jupiter mass primary (Chauvin et al. 2005; Skemer et al. 2011). 2M1207b likely formed by fragmentation and so did not form from a disk, similar to planetary-mass object OTS44 which has Paschen-β emission (Joergens et al. 2013). All other known "planetary mass accreting companions" also likely formed as brown dwarfs through gravitational instability and/or fragmentation.

   Enigmatic "resolved" (by binary sparse aperture models) companion planet candidates such as LkCa Hα 15b and T Cha b (Kraus & Ireland 2012; Huélamo et al. 2011) lie in the gap regions of transitional disks. More work is needed to establish if these interesting objects are real (Olofsson et al. 2013) and actively accreting. In this letter we will, for the first time, characterize the Hα accretion of a stellar companion, well inside a transitional disk gap, and introduce a survey to detect planetary mass objects using the same methodology.

*1.1.1 How Can We Detect Accreting Gas Giants?*

A survey of young disks at Hα would be very sensitive to low-mass planets if they have even moderate accretion rates ($M_{dot} > 10^{-10}$ $M_{sun}$/yr). With the novel MagAO SDI mode (see section 1.2.1) we can obtain *simultaneous* PSF and photometric calibration through the 642 nm continuum image. Using continuum scaling from non-accreting flat-spectrum (unpolarized) calibration stars, simultaneous acquisition of the continuum allows for robust PSF subtraction, isolating any Hα point sources and allowing us to quantify the amount of accretion through Hα excess.

*1.1.2 Will Extinction Absorb all the Hα Emission?*

In the case of most young stars, extinction due to disk material would preclude us from directly observing the midplane, where accreting objects are likely to lie. However, transitional disks host large 1-100+ AU "gaps" that are fully or partially cleared of dusty material, making them optically thin. In fact, it has been theorized that one reason these gaps may stay open is the sculpting influence of giant planets (Alexander & Armitage 2009; Owen & Clarke 2012). Perhaps the best candidates are the so-called "wide gap" transitional disks. These 10-100 AU gaps are many times the size of any one planet's Hill sphere $(M_p/3M_*)^{1/3}r$, and even a massive 5 $M_{jup}$ planet can only open an ~6 AU gap at r=10 AU. Therefore, these disks likely need multiple >1 $M_{jup}$ mass planets to keep the gap cleared. The simulations of Dodson-Robinson & Salyk (2011) make a convincing case that multiple (2-3) massive (1-10 $M_{jup}$) planets can reproduce all (five) of the main properties observed in transitional disks (see fig. 1). If this is indeed the scenario through which these gaps are formed, then one should be able to directly detect Hα accretion onto the planet/companion (or onto a shock surface/boundary) in transitional disks of moderate inclination without prohibitively high extinction along the line of sight through the gap.

**1.2 MagAO: A New Level of Wavefront Control**

The main issue preventing high contrast imaging at Hα is the lack of a large telescope AO system that can achieve even modest (>20%) Strehls at the "blue" wavelength of Hα (0.656 μm). However, a new 585 actuator 1000 Hz AO System (MagAO) has been commissioned (since May 2013) at the 6.5m Magellan Clay Telescope. Magellan's site is one of the best seeing locations in the world, with a median seeing of 0.64". The 378 corrected modes of the MagAO system, makes it the highest sampled (a ~23 cm actuator pitch) AO system on a large (>5m) telescope today. It feeds both a visible (VisAO) and infrared camera (Clio2) simultaneously. Indeed the first light results of Close et al. (2013) prove that MagAO can produce very high spatial resolutions (20-30 mas) in the visible (as short as $\lambda=0.6\mu m$) for the first time.

*1.2.1. The VisAO Camera's SDI Mode: Differential Hα Imaging*

To achieve high-contrast, even with AO, simultaneous PSF information to compare to (or deconvolve against) the "in-line" science image is necessary. An extremely effective technique for obtaining a simultaneous PSF is Simultaneous Differential Imaging (SDI; Marois et al. 2003; Close et al. 2005), which utilizes a Wollaston Beamsplitter to obtain nearly identical simultaneous, images of the *o-polarized* and *e-polarized* beam, yet some residual errors remain from non-linear chromatic variation of speckles and non-common path (NCP) errors.

The SDI configuration of Magellan's VisAO camera includes: 1) a thin, small-angle, calcite Wollaston beamsplitter near the pupil; and 2) a split on-Hα/off Hα "SDI" filter just before the focal plane (for the *e* beam/*o* beam). In this mode, we obtain a good (photon noise limited) simultaneous calibration of the PSF "off" and "on" the Hα line. Hence, a simple subtraction of the "off" image from the "on" image (with a simple throughput scaling) will map Hα structures (jets, disks, accreting faint companions etc.), with minimal photometric confusion from the continuum or structural confusion from the PSF (except, unfortunately, for NCP narrow-band filter "ghosts" that are different in each filter and so don't subtract). For examples of the power of this observing mode, see the Orion Silhouette disk resolved with this Hα SDI mode of MagAO (Follette et al. 2013) or the LV1 binary Proplyd imaged at Hα with MagAO (Wu et al. 2013). Coupling continuum SDI subtraction with a standard ADI (Angular Differential Imaging; Marois et al. 2006) reduction produces ASDI very high-contrast Hα images.

## 2.0 OBSERVATIONS & REDUCTIONS

### 2.1. Hα SDI observations of the HD142527 Transitional Disk System

The potential of this Hα SDI mode for detecting accreting objects in the gaps of transitional disks encouraged us to take a "test" survey observation of the young (5 Myr) Sco OB-2 association F6IIIe HerbigAeBe star HD 142527, which has a large (r~140AU) gap in its transitional disk. HD 142527 is an excellent example of the type of transitional disk where directly imaging accreting planets at Hα may be possible. In particular, it has a large, optically-thin, gap (Fukagawa et al. 2006) stretching from ~10 AU to ~140 AU that is nearly face-on ($i$~20$^o$, Casassus et al. 2013). Moreover, there is significant stellar accretion (~7x10$^{-8}$ $M_{sun}$/yr; estimated by Garcia Lopez et al. 2006) occurring, which requires gas to stream into the gap (perhaps in streamers seen by ALMA;

Casassus et al. 2013) to prevent the inner gas disk from being consumed in less than ~1 yr. Multiple giant "gap planets" in this system (see Fig. 1) would both allow "streamers" to move gas through the huge gap and also allow the majority gap to stay optically thin as debris is cleared/scattered by planets in the inner and outer parts of the disk (Owen & Clarke 2012). Gas will "seek out" any gravitational potential wells (planets) on its way to the star, and hot gas in these wells (planets) should emit in Hα.

Recently Biller et al. (2012) observed HD 142527 with Sparse Aperture Mask (SAM) interferometry with VLT/NACO at H, K and L' (3.8 microns) and found a stellar companion candidate at 88±4 mas separation and 133±3° PA. This candidate was not detected in any of the other surveys of this object, and Casassus et al. (2013) cast doubt on this companion's existence since an asymmetric inner disk might mimic such a binary detection.

During the second (and final) commissioning run of the MagAO system, we observed HD142527 on 2013 April 11 (UT). We made Hα SDI observations in ADI mode. The night was photometric, and V-band seeing varied from 0.4" to 0.9" (0.64" is the site median) during our observations (see Fig. 2). Over 150 degrees of ADI sky rotation and 1.9 hours of "best" (Strehl>19%) open shutter time (2,950 short unsaturated 2.3s exposures).

*2.1.1. The First SDI Hα imaging of a Transitional Disk: An Accreting Companion Found*

The data was reduced in usual fashion (Close et al. 2013) with sky/bias subtraction and then cross-correlation alignment. The CCD frames are flat to <1%, and the SDI platescale is 0.00798±0.00002"/pix (see Close et al. 2013 for calibration details). Figures 3a and 3b show the full standard ADI reduction (Marois et al 2006) of all 2,950 Hα frames with wavefront control better than 132 nm rms (Hα Strehl ≥ 19%). Fig 3c shows the ASDI reduction. ASDI reduction is an ADI reduction of "Hα-minus-Continuum" residual images. As is clear from Figs 3b and 3c, there is a Hα source at sep=86.3±1.9 mas, PA=126.6±1.4$^o$. Astrometric errors from the average of ADI and ASDI sources detected by PSF *allstar* fitting photometry. The best photometry for B avoided self-subtraction ADI errors by use of "fake companions" which were subtracted from six different 200 image sets of the rotated aligned data near the location of B and iterated in flux and position until the flux from B was subtracted away from the Hα and Cont. images (see Close et al. 2007 for more details). The best range of B positions and relative photometry (ranging from under-subtracted to over-subtracted) found, over the six images, were sep=84.2±2.5 mas; PA=127±2$^o$; ΔmagHα=6.33±0.20 mag; ΔmagCont=7.50±0.25 mag.

# 3.0 DISCUSSION

Clearly there is a detection of a point-source (FWHM$_B$~26 mas almost identical to the PSF FWHM$_A$~25 mas) very close to the position of the candidate of Biller et al. (green circle) in Fig. 3. Our position is >7σ inconsistent with a background object at our epoch (61±4 mas, PA=117±3$^o$; white circle Fig 3). Hence, this is a real physical companion, henceforth HD142527B. Comparing our position to Biller et al. suggests 6±4$^o$/yr of clockwise motion. This is consistent with a circular orbit about the ~2.5 M$_{sun}$ primary with a semi-major axis (de-projected) of ~14 AU, however more eccentric orbits can also fit the sparse data. No significant extended Hα structures/streamers (such as those of Casassus et al. 2013 in the sub-mm) were detected, other than a very weak outline of the outer 1" radius "ring" seen in the NIR. Longer exposures than the 2.3s used here are required to probe extended outer structures at Hα.

The Hα emission from B is ~300% that of the continuum. Our detection of a continuum source 1,000x fainter than the primary just 86 mas away is a new contrast record for astronomy and speaks to the power of moderate Strehls in the visible (plus ADI/ASDI). While this contrast is noteworthy, it is the exciting discovery of an actively accreting source at a projected separation of 12 AU – right at the inner edge of the HD142527 disk (~10 AU; Casassus et al. 2013) that is most important. It is reasonable to assume, based on its location, that HD142527B is responsible for clearing/maintaining the inner ~10 AU gap-edge in the HD 142527 transitional disk. If on a very eccentric orbit it could be responsible for the entire gap (Biller et al. 2012).

**3.1 What Can HD142527B Tell Us About Accreting Objects in Transitional Disks?**

Our detection of an accreting low-mass companion in Fig. 3 opens an exciting new window on the coupled processes of accretion onto a circumsecondary disk and accretion from the circumsecondary disk to the forming planet (or brown dwarf/low-mass star). While long envisioned and modeled (e.g. Lubow et al. 1999; Gressel et al. (2013) or d'Angelo & Bodenheimer (2013)), these processes have never been observed at the level of our observations. Models predict that a massive gas giant planet will open a gap, but can continue to accrete from the circumstellar disk through spiral-arm-like gas streamers. Although we detect a low-mass stellar companion here, the accretion steamers could be similar. High-angular-momentum gas, however, cannot directly land on the compact object but will instead form a small "circumsecondary" disk. This circumsecondary disk will play a similar role to accretion disks around stars, but its physical and thermal structure will be more complex (e.g. Martin & Lubow 2011). Indeed Biller et al. observe a K-L'=0.9±0.3 for HD142527B which is a sign of ~0.3 excess NIR flux from a circumsecondary disk around HD142527B.

Hα emission from a circumsecondary disk has not been studied theoretically yet, but it is likely that the geometry of the shocked region and the accretion column will be similar to those for accreting brown dwarfs/low mass stars. In addition, an accretion shock may also form where the streamers from the circumsecondary disk deliver gas to the denser circumsecondary disk.

**3.2 Do these Objects Really Have a Large Equivalent Width in Hα?**

Although the theoretical framework to fully interpret Hα emission from accretion through a circumsecondary disk is currently lacking, we attempt a simple comparison of HD 142527B to accreting brown dwarfs and low-mass stars observed in the Cha I star-forming region (age 1-3 Myr), based on a sample discussed in Szucs et al. (2010) and using Hα measurements from Nguyen et al. (2009, 2011). In Fig. 4 we compare the Hα equivalent widths and approximate luminosities to the source detected around HD 142527 (the $L_{bol}$ plotted in fig 4 is derived in the next section). Here we estimated the equivalent width by EW=dλ x $F_{Hα}/F_{cont}$, where dλ is the width of the 643 nm continuum filter (0.006 microns), $F_{Hα}$ and $F_{cont}$ are the flux densities measured in the Hα and continuum filters, respectively. This exploratory comparison highlights that HD 142527B occupies a different parameter space than isolated accreting young brown dwarfs/low mass stars.

We speculate that this strong Hα emission relative to the continuum is an indicator of strong accretion, perhaps reflecting the fact that the overall gravitational potential gas reservoir (which determines the gas heating) is dominated by the primary T-Tauri star, while the local continuum emission is dominated by the companion's photosphere. Hence the luminosity in Hα is very high even though the mass of the companion may be very low. We derive just how high this Hα luminosity can be in the next sections.

**3.3 Extinction**

It appears that this object has considerable excess dust emission (H=10.5; H-Ks=0.5; Ks-L'=0.9; Biller et al. 2012) these very red colors are only consistent with an optically thick circumsecondary disk around HD142527B itself. Biller et al. 2012 find for B $M_{Hobs}$=4.7±0.3. Assuming that the disk of HD142527B creates a small ~0.3 mag disk H excess ($H_{excess}$; as estimated from Biller et al. NIR photometry) then the true photospheric H is $H_{true}=H_{obs}+0.3-A_H$. Assuming a small $A_H$~0.3mag extinction cancels the small 0.3mag disk excess at H, then $M_{Htrue}$=4.7±0.3. The 5Myr isochrones of Baraffe et al. (1998) suggests that B (at $M_{Htrue}$=4.7±0.3) is a 0.25 $M_{sun}$ object in agreement the masses of Biller et al. 2012. The same isochrone yields $M_R$~8.7 mag at that mass. Since the R magnitude for component A is $R_A$=8.3±0.1 mag and since ΔmagCont=7.50±0.25 mag then $R_B$=15.8±0.3. Then $M_R$=10±0.3 is the absolute mag for B (D=145±5pc; Biller et al. 2012). Hence the R band (and Hα) of B is extincted by $A_R$=10-8.7=1.3±0.3 mag (similar to the low extinction observed towards A; Malfait et al. 1999). As a sanity check, our independent extinctions of $A_R$=1.3±0.3 mag and $A_H$~0.3 mag ($A_H/A_R$= 0.23±0.07) are in good agreement with the dust law of Cardelli et. al (1989) where we expect $A_H/A_R$=0.22.

**3.4 Hα Line Luminosity**

Log($L_{Hα}$) luminosity can be calculated from:

Log[(4πD$^2$)Vega_0_ Hα _c*Filter_width /($10^{((R_A+\Delta magH\alpha-A_R)/2.5)}$) /$L_{sun}$] which here is

= log[4π(145*3.1x10$^{18}$)$^2$*2.339x10$^{-5}$*0.006/($10^{((8.3+6.3-1.3)/2.5)}$)/3.8x10$^{33}$] = -3.3 $L_{sun}$ where the Vega zero magnitude in our Hα filter is calculated to be 2.339x10$^{-5}$ ergs/(s cm$^2$ μm).

### 3.5 Accretion Luminosity and the Mass Accretion Rate

Assuming the TTS relationship in which log ($L_{acc}$) = b + a * log ($L_{Hα}$) holds, then from Rigliaco et al (2012) $L_{acc} = 10^{[2.27\pm0.23+ (1.25\pm0.07)*(-3.3)]}$ ~ 1.3% $L_{sun}$, a substantial accretion luminosity for a very low mass object. This allows us to calculate the Mass accretion rate as $M_{dot\_companion}$ = 1.25[$L_{acc}$]$R_{comp}$/(G$M_{comp}$) (Gullbring et al. 1998), yielding 5.9x$10^{-10}$ $M_{sun}$/yr. We can assume that this is a reasonable estimate for the accretion flow for other planets in the HD142527 gap as well (Dodson-Robinson & Salyk 2011). We note this is a very reasonable planetary accretion rate (adding ~1.2 Mjup of planet mass in ~2 Myr). Compared to the estimated accretion rate of the star ($M_{dot}$~7x$10^{-8}$ $M_{sun}$/yr), this calculation suggests that only ~1% of the gas falls onto HD142527B and presumably any other planets in the gap. Theory (Lubow et al 1999) actually predicts planetary accretion rates can be even ~10x higher than this, so maybe this is a quiescent period for accretion on HD142527B, or maybe this accretion rate is reasonable for B's mass. Also the accretion may have been overestimated by Garcia Lopez et al. (2006) by a factor of ~2. In any case, we can assume that this $M_{dot\_companion}$ value holds (as a minimum) throughout the gap and reverse the steps above to derive the $L_{acc}$, $L_{Hα}$, and finally ΔmagHα for lower mass (planetary) companions from the formulas above. This is illustrated in Fig. 5 where the planet/star contrast at Hα is estimated to be better (higher values) than at any other wavelength (even with 3.4 mag of extinction) for low-mass (<5 $M_{jup}$) accreting "gap" planets. This opens an exciting new observational window for the direct detection of accreting low mass ~0.5-5 $M_{jup}$ planets (that would otherwise be very hard to directly detect in the NIR).

# 4.0 FUTURE OBSERVATIONS

## 4.1 GAPplanetS: a Survey for Hα Planets in the Gaps of Transitional Disks

We have started the **G**iant **A**ccreting **P**roto-**planets S**urvey (GAPplanetS). This is a survey of young (~5 Myr) nearby (D ≤ 250pc) transitional disks with SDI Hα (and simultaneous Brγ imaging with MagAO's Clio IR camera) for all known transitional disks observable with MagAO. From our "pilot" HD142527 observations we know that in ≤ 0.8" seeing and ≤ 20 mph winds we have the required Hα Strehls and contrasts (Fig. 2). **Hence, on >75% of the nights at LCO MagAO should obtain the quality of correction needed for ≥ 20% Strehls at Hα** (Thomas-Osip et al. 2008).

The contrasts ($10^{-4}$) already achieved in the Hα ADI image in Fig. 3 are more than adequate to detect 1 Jupiter mass planets ($\dot{M}_{companion}$=6x$10^{-10} M_{sun}$/yr; $A_R$~0mag) at 5σ and sep ≥ 0.2" according to Fig 5. While there are some unknowns in these estimates (accretion rate, extinction), it is clear that GAPplanetS has great potential to be very sensitive to lower mass (~1 $M_{jup}$) accreting giant planets in transitional disks.

We thank the anonymous referee for helpful suggestions. MagAO was constructed with NSF MRI, TSIP, and ATI awards.

**REFERENCES**
Alexander, Richard D.; Armitage, Philip J. 2009 ApJ, 704, 989
Apai, D. et al. 2004, A&A 415, 671
Baraffe, I.; Chabrier, G.; Allard, F.; Hauschildt, P. H. 1998, A&A 337 403
Baraffe, I.; Chabrier, G.; Barman, T.S. et al. 2003, A&A 402 701
Biller, Beth; Lacour, Sylvestre; Juhász, Attila; Benisty, Myriam; Chauvin, Gael; Olofsson, Johan; Pott, Jörg-Uwe; Müller, André; Sicilia-Aguilar, Aurora; Bonnefoy, Mickaël 2012 ApJ 753 38
Bonnefoy, M.; Boccaletti, A.; Lagrange, A.-M.; Allard, F.; Mordasini, C.; Beust, H.; Chauvin, G.; Girard, J. H. V.; Homeier, D.; Apai, D.; 2013 A&A 555, A107
Boss, A. P. 2006, ApJ, 643, 501
Cardelli, Clayton, and Mathis 1989 ApJ 345 245
Casassus, Simon; van der Plas, Gerrit; M, Sebastian Perez; Dent, William R. F.; Fomalont, Ed; Hagelberg, Janis; Hales, Antonio; Jordán, Andrés; Mawet, Dimitri; Ménard, Francois 2013, Nature 493, 191
Chauvin, G., Lagrange, A.-M., Dumas, C., Zuckerman, B., Mouillet, D., Song, I., Beuzit, J.-L., Lowrance, P. 2005, A&A, 438, 25.
Close, L.M. et al. 2005 Nature 433, 286.
Close L.M. et al. ApJ 2007, ApJ, 660, 1492

# High-Contrast Imaging of the Accreting Companion HD142527B at Hα with MagAO

Close, L.M. et al. 2013, ApJ, 774, 94.
D'Angelo, G.; Bodenheimer, P. 2013, ApJ 778, 77
Dodson-Robinson S, & Salyk C. 2011 ApJ, 738, 131
Follette, K. et al., 2013, ApJ, 775, 13
Fortney, J., Marley, M., Saumon, D., Lodders, K. 2008, ApJ 683, 1104
Fukagawa, Misato; Tamura, Motohide; Itoh, Yoichi; Kudo, Tomoyuki; Imaeda, Yusuke; Oasa, Yumiko; Hayashi, Saeko S.; Hayashi, Masahiko 2006, ApJ Lett 636, L153
Garcia Lopez, R., Natta, A., Testi, L., & Habart, E. 2006, *A&A*, **459**, 837
Gullbring, Erik; Hartmann, Lee; Briceno, Cesar; Calvet, Nuria 1998, ApJ 492, 323
Gressel, O.; Nelson, R. P.; Turner, N. J.; Ziegler, U. 2013 ApJ 779 59
Huélamo, N.; Lacour, S.; Tuthill, P.; Ireland, M.; Kraus, A.; Chauvin, G. 2011, A&A 528, 7.
Jayawardhana, R., Ivanov V.D. 2006, Science, 313 1279
Joergens, V.; Bonnefoy, M.; Liu, Y.; Bayo, A.; Wolf, S; Chauvin, G.; Rojo, P. 2013, A&A 558, 7
Kraus, Adam L.; Ireland, Michael J. 2012, ApJ 745, 5
Lagrange, A.-M.; Gratadour, D.; Chauvin, G.; Fusco, T.; Ehrenreich, D.; Mouillet, D.; Rousset, G.; Rouan, D.; Allard, F.; Gendron, É.; Charton, J.; Mugnier, L.; Rabou, P.; Montri, J.; Lacombe, F. 2009 A&A 493, 21
Liu, M. C., Leggett, S. K., Golimowski, D. A., Chiu, K., Fan, X., Geballe, T. R., Schneider, D. P., & Brinkmann, J. 2006, ApJ, 647, 1393
Lubow et al. 1999 ApJ 526, 1001
Luhman, K.L. et al. 2007, ApJ 659, 1629
Malfait, K., Waelkens, C., Bouwman, J., de Koter, A., & Waters, L. B. F. 1999 A&A, 345, 181
Marois, C., Nadeau, D., Doyon, R., Racine, R., & Walker, G. A. H. 2003, in IAU Symposium, 275
Marois, C. et al. 2006 ApJ, 641 556
Marois C. et al. 2008, Science 322, 1348
Martin & Lubow 2011 MNRAS 413, 1447
Mohanty, S. et al. 2007 ApJ 657, 1064
Mordasini, C.; Alibert, Y.; Klahr, H.; Henning, T. 2012, A&A 547, 111
Muzerolle, James; Hillenbrand, Lynne; Calvet, Nuria; Briceño, César; Hartmann, Lee 2003 ApJ 592 266
Nguyen, Duy Cuong; Scholz, Alexander; van Kerkwijk, Marten H.; Jayawardhana, Ray; Brandeker, Alexis, 2009, ApJ 694, 153
Olofsson, J.; Benisty, M.; Le Bouquin, J.-B.; Berger, J.-P.; Lacour, S.; Ménard, F.; Henning, Th.; Crida, A.; Burtscher, L.; Meeus, G.; Ratzka, T.; Pinte, C.; Augereau, J.-C.; Malbet, F.; Lazareff, B.; Traub, W. 2013 A&A 552, 4
Owen, J.E. & Clarke, C.J. 2012, MNRAS 426, 96
Rigliaco et al. 2012, A&A 548, A56
Skemer, Andrew J.; Close, Laird M.; Szűcs, László; Apai, Dániel; Pascucci, Ilaria; Biller, Beth A. 2011, ApJ 730 53
Skemer, Andrew J. et al. 2012, ApJ 753 14
Spiegel, D.S. & Burrows A. 2012, ApJ 745, 174
Thomas-Osip, J. E.; Lederer, S. M.; Osip, D. J.; Vilas, F.; Domingue, D.; Jarvis, K.; Leeds, S. L. 2008 proc SPIE Volume 7012, article id. 70121U
Szűcs, László; Apai, Dániel; Pascucci, Ilaria; Dullemond, Cornelis P. 2010, ApJ 720 1668
Wu, Y-L. et al. 2013, ApJ 774, 45

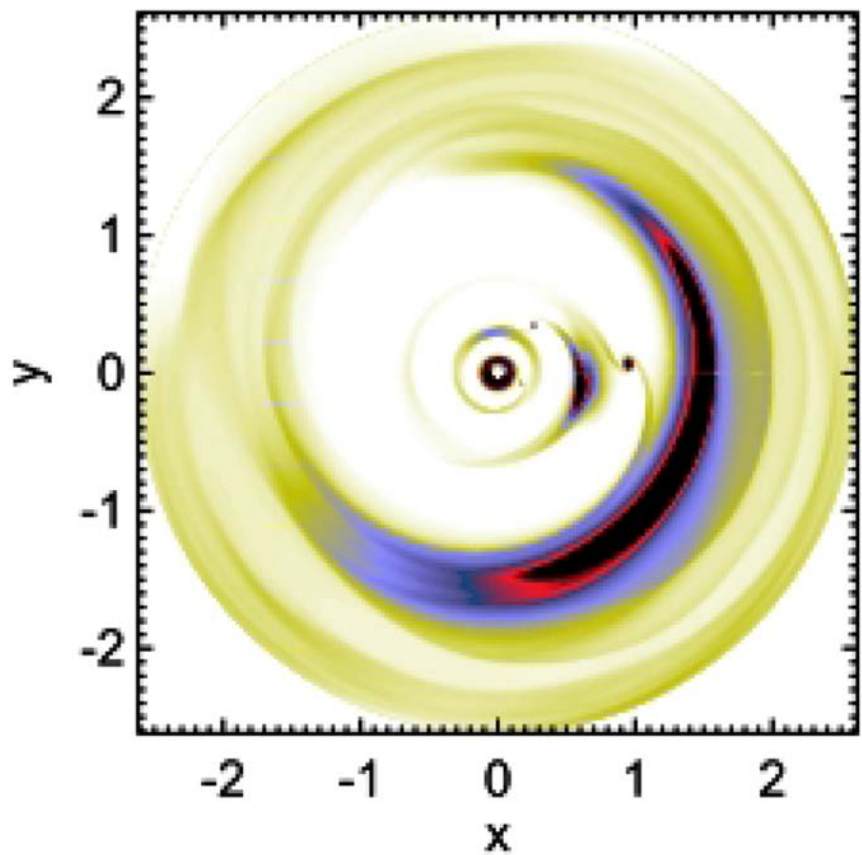

**Fig 1:** Simulations of how a multiplanet system (3 in this case) can maintain a wide gap in a transitional disk (reproduced from Dodson-Robinson & Salyk 2011).

**High-Contrast Imaging of the Accreting Companion HD142527B at Hα with MagAO**

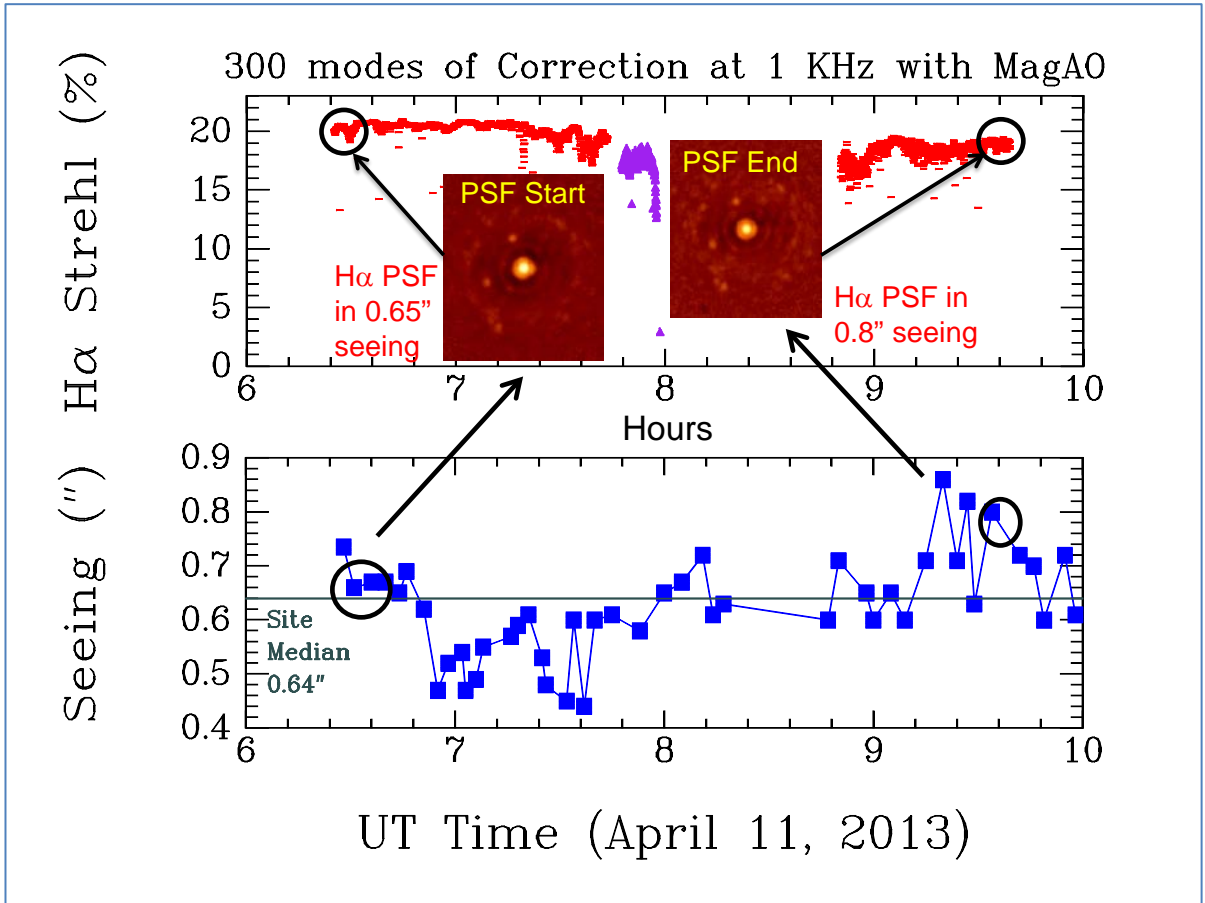

**Fig 2:** Over our ~2.5 hour (clock time) observation of HD 142527 (3,256 2.3s images) moderate Strelhs were obtained in typical conditions. *Insets* deep log Stretch of the median of the first 200 images and the last 200 images, in worse seeing, after 150 degrees of rotation, and on a different part of the CCD. These are almost identical images which are excellent for achieving high-contrasts with ADI.

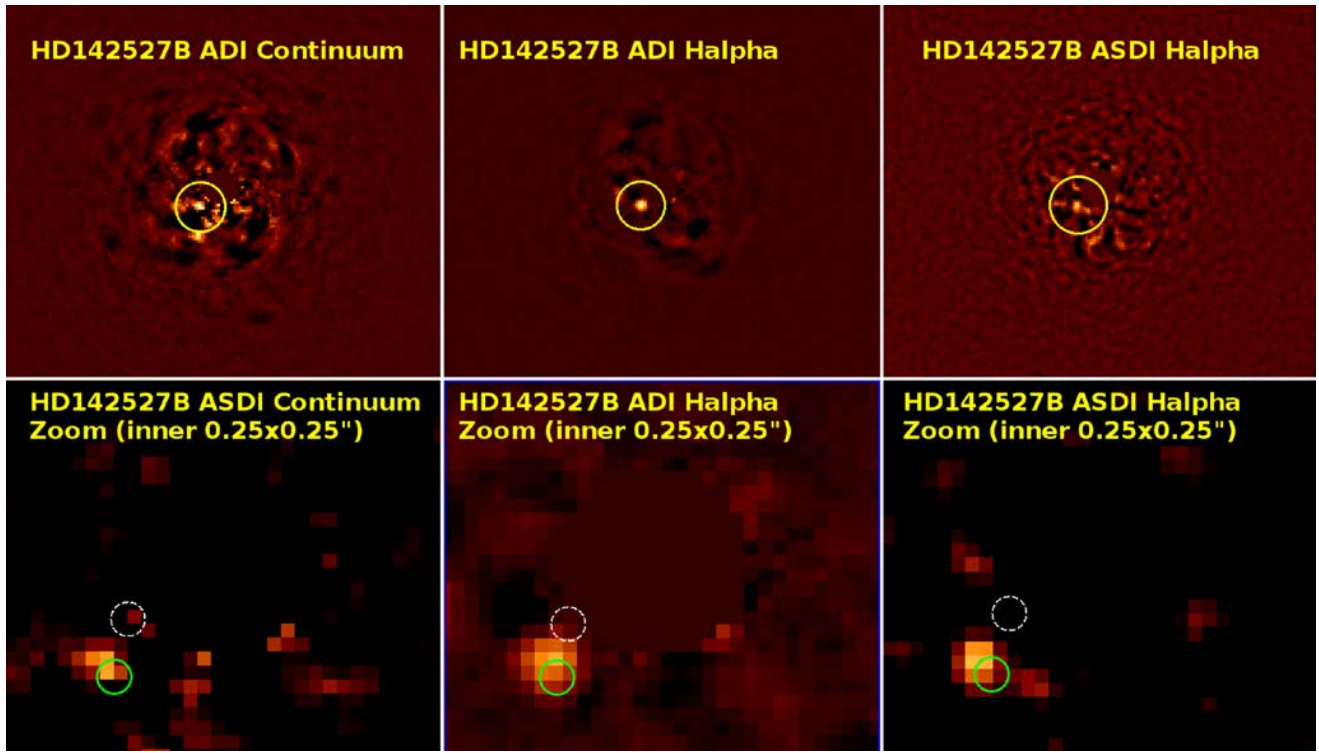

**Fig 3a:** (*left*) Continuum (643 nm) ADI reduced image. (*Bottom Zooms*): Note the weak (~3 sigma) detection near the location of the candidate of Biller et al. 2012 (green circle). The source is inconsistent with the background star position (white circle). **3b:** (*middle*) Hα ADI images. Note the unambiguous 10.5 sigma Hα point-source at sep=86.3 mas, PA=126.6$^o$, hereafter HD142527B. **3c:** (*right*) ASDI data reduction, here NCP narrow-band filter ghosts are not as well removed as with ADI.

**High-Contrast Imaging of the Accreting Companion HD142527B at Hα with MagAO**

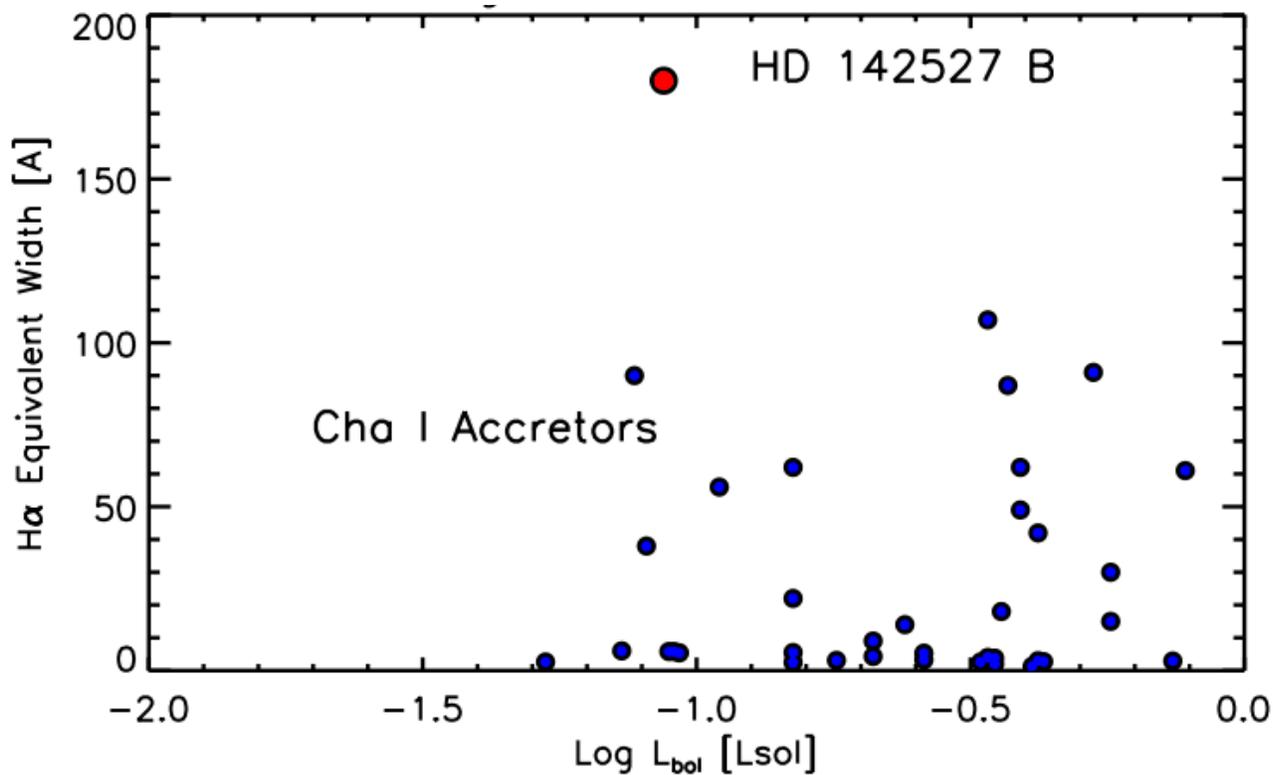

**Fig. 4:** The companion (HD 142527B – red dot), inside the gap of a transitional disk, has significant accretion emission compared to younger low mass stars and brown dwarfs.

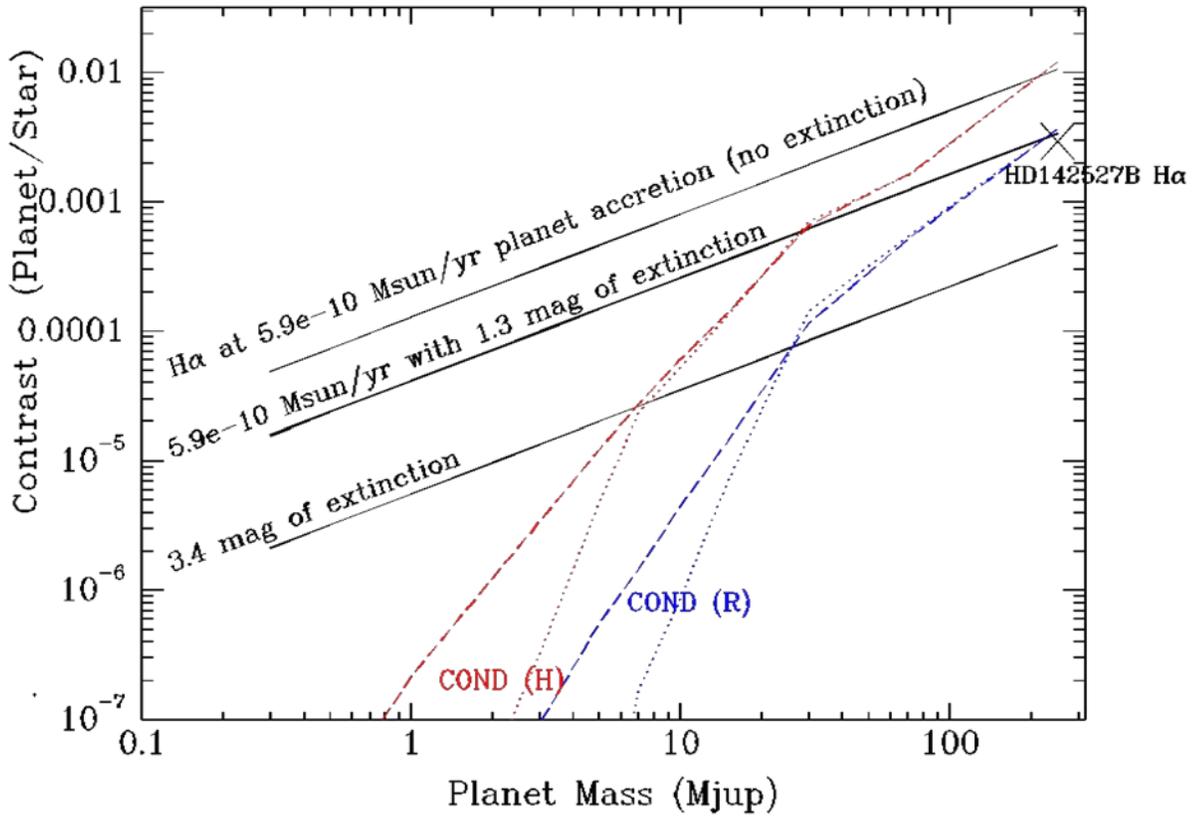

**Fig. 5:** Here we estimate how bright planets can appear in Hα (dark straight lines) when there is 5.9x10$^{-10}$ M$_{sun}$/yr accreting on them. Depending on how much extinction there is, a 1 Jupiter mass planet can have an Hα contrast that is 50-1,000x better (A$_R$=3.4-0 mag) than the most optimistic (5Myr COND; Baraffe et al. 2003) atmospheric models (dashed red line H band, blue dashed line R band). The 5 Myr DUSTY models are plotted as dotted lines. The radius of all young brown dwarfs and planets are assumed to be similar.